\documentclass[pageno]{jpaper}

\usepackage{mathptmx} %

\usepackage{fancyhdr}

\usepackage{upgreek} %
\usepackage{caption,subfig}
\usepackage{graphicx}
\usepackage{xcolor}
\usepackage{algorithmic}
\usepackage{xspace}
\usepackage{multirow}
\usepackage{array}
\usepackage{tikz}
\usepackage{balance}
\usepackage{enumitem}
\usepackage{comment}
\usepackage{soul}
\usepackage{todonotes} %
\usepackage{tcolorbox}
\usepackage{wrapfig}
\usepackage{rotating}
\usepackage{stfloats} %
\usepackage{wrapfig}

\usepackage[capitalize,noabbrev,nameinlink]{cleveref}
\crefformat{section}{#2§#1#3}
\crefformat{subsection}{#2\S#1#3}
\crefformat{subsubsection}{#2\S#1#3}

\definecolor{iris}{rgb}{0.35, 0.31, 0.81}

\newcommand{\TheName}{\mbox{\textsc{CoaXiaL}}\xspace}
\newcommand{\simulator}{\mbox{ChampSim}\xspace}

\Crefname{figure}{Fig.}{Figs.}
\crefname{figure}{Fig.}{Figs.}

\newif\ifvariance
\variancetrue

\def\BibTeX{{\rm B\kern-.05em{\sc i\kern-.025em b}\kern-.08em
    T\kern-.1667em\lower.7ex\hbox{E}\kern-.125emX}}

\title{A Case for CXL-Centric Server Processors}

\author{Albert Cho$^*$\quad\quad Anish Saxena$^*$\quad\quad Moinuddin Qureshi\quad\quad Alexandros Daglis\\ Georgia Institute of Technology\vspace{-3mm}}
{
   \fancyhf[C]{\Large Work under submission -- Please do not distribute!}
   \fancyfoot[C]{\thepage}
}

\begin{document}
\date{}
\maketitle

\begin{abstract}

The memory system is a major performance determinant for server processors. 
Ever-growing core counts and datasets demand higher bandwidth and capacity as well as lower latency from the memory system.
To keep up with growing demands, DDR---the dominant processor interface to memory over the past two decades---has offered higher bandwidth with every generation.
However, because each parallel DDR interface requires a large number of on-chip pins, the processor's memory bandwidth is ultimately restrained by its pin-count, which is a scarce resource.
With limited bandwidth, multiple memory requests typically contend for each memory channel, resulting in significant queuing delays that often overshadow DRAM's service time and degrade performance.

We present \TheName, a server design that overcomes memory bandwidth limitations by replacing \textit{all} DDR interfaces to the processor with the more pin-efficient CXL interface. 
The widespread adoption and industrial momentum of CXL makes such a transition possible, offering $4\times$ higher bandwidth per pin compared to DDR at a modest latency overhead.
We demonstrate that, for a broad range of workloads, CXL's latency premium is more than offset by its higher bandwidth.
As \TheName distributes memory requests across more channels, it drastically reduces queuing delays and thereby both the average value and variance of memory access latency.
Our evaluation with a variety of workloads shows that \TheName improves the performance of manycore throughput-oriented servers by $1.52\times$ on average and by up to $3\times$.

\end{abstract}

\pagestyle{plain}

\sloppy
\let\thefootnote\relax\footnote{$^*$Equal contribution.}
\section{Introduction}
\label{sec:intro}

Multicore %
processor architectures have been delivering performance gains despite the end of Dennard scaling and the slowdown of Moore's law in the past two decades. 
At the same time, as the data consumed by processors is increasing exponentially, technological breakthroughs have enabled higher-capacity memory with new media like non-volatile RAM or via remote memory access over fast networks (e.g., RDMA). 
A common technological trade-off with higher-capacity memory is significantly inferior memory access latency and bandwidth compared to the DDR-based main memory. 
As a result, servers continue to predominantly rely on DDR-attached memory for performance while optionally retaining a slower memory tier like NVRAM or remote DRAM for capacity expansion.

The emerging Compute Express Link (CXL) standard bridges the performance gap between low-bandwidth, high-capacity memory and DDR-based main memory. 
By attaching DRAM modules over the widely deployed high-bandwidth PCI Express (PCIe) bus, CXL vastly improves memory capacity and bandwidth, while retaining DDR-like characteristics at a modest access latency overhead. 
Consequently, there has recently been much interest in architecting CXL-based memory systems that enable memory pooling and capacity expansion \cite{ahn:enabling,gouk:direct,li:pond,marouf:tpp}. %

CXL owes its high bandwidth to the underlying PCIe-based serial interface, which currently delivers about $4\times$ higher bandwidth per processor pin compared to the parallel DDR interface, with technological roadmaps projecting this gap to grow further. 
Hence, by repurposing the processor's DDR-allocated pins to CXL, it is possible to quadruple the available memory bandwidth.
However, the higher bandwidth comes at the cost of memory access latency overhead, expected to be as low as 25--30ns \cite{cxl-3.0, plda:cxl-latency}, although higher in initial implementations and systems that multiplex CXL memory devices across multiple processors. 
Low access latency is a key requirement for high-performance memory,  which is why CXL's latency overhead has biased the research so far to treat the technology exclusively as a memory \textit{expansion} technique rather than a \textit{replacement} of local DDR-attached memory.

However, we observe that the overall memory access latency in a loaded system is dominated by the queuing delay at the memory controller, which arbitrates access to the DDR channel. 
Modern servers feature between 4 and 12 cores per memory channel, resulting in contention and significant queuing delays even before a request can be launched over the memory bus. 
Mitigating these queuing delays by provisioning more memory channels requires more processor pins and die area, which are scarce resources. %
Given rigid pin constraints, CXL's bandwidth-per-pin advantage can unlock significant bandwidth and performance gains by rethinking memory systems to be CXL-centric rather than DDR-centric. %

In this paper, we make the key observation that the bandwidth boost attainable with  CXL drastically reduces memory access queuing delays, which largely dictate the effective access latency of loaded memory systems. 
In addition to increased average memory access latency, queuing delays also increase memory access variance, which we show has detrimental effects to performance.
Driven by this insight, we argue that \textit{a memory system attached to the processor {entirely} over CXL} is a key enabler for scalable high-performance server processors that deploy memory-intensive workloads. %
Our proposed server design, dubbed \TheName, replaces all of the processor's direct DDR interfaces with CXL.

By evaluating \TheName with a wide range of workloads, we highlight how CXL-based memory system's unique characteristics (i.e., increased bandwidth and higher unloaded latency) positively impact performance of processors whose memory system is typically loaded.
Our analysis relies on a simple but often overlooked fact about memory system behavior and its impact on overall performance:
that a loaded memory system's effective latency is dominated by the impact of queuing effects and therefore significantly differs from the unloaded system's latency, as we demonstrate in \cref{sec:motivation:queuing-impact}. 
A memory system that offers higher parallelism reduces queuing effects, which in turn results in lower average latency and variance, even if its unloaded access latency is higher compared to existing systems.
We argue that CXL-based memory systems offer exactly this design trade-off, which is favorable for loaded server processors handling memory-intensive applications, offering strong motivation for a radical change in memory system design that departs from two decades of DDR and enables scalable high-performance server architectures.

In summary, we make the following contributions:
\begin{itemize}[noitemsep,topsep=0pt,leftmargin=*]

    \item We make the radical proposal of using high-bandwidth CXL as a \textit{complete replacement} of pin-inefficient DDR interfaces on server processors, showcasing a ground-breaking shift that disrupts decades-long memory system design practices. 
    \item We show that, despite its higher unloaded memory access latency, \TheName reduces the effective memory access time in typical scenarios where the memory system is loaded.
    \item We demonstrate the promise of \TheName with a study of a wide range of workloads for various CXL bandwidth and latency design points that are likely in the near future. 
    \item We identify limitations imposed on CXL by the current PCIe standard, and highlight opportunities a revised standard could leverage for 20\% additional speedup.
\end{itemize}

\smallskip
\noindent\textit{Paper outline:}
\cref{sec:motivation} motivates the replacement of DDR with CXL in server processors.
\cref{sec:queuing-effect} highlights the critical impact of queuing delays on a memory system's performance and \cref{sec:design} provides an overview of our proposed \TheName server design, which leverages CXL to mitigate detrimental queuing.
We outline the methodology to evaluate \TheName against a DDR-based system in \cref{sec:method} and analyze performance results in \cref{sec:eval}.
We discuss related work in \cref{sec:related} and conclude in \cref{sec:conclusion}.

\section{Background}
\label{sec:motivation}

In this section, we highlight how DRAM memory bandwidth is bottlenecked by the processor-attached DDR interface and processor pin-count. 
We then discuss how CXL can bridge this gap by using PCIe as its underlying physical layer. %

\subsection{Low-latency DDR-based Memory }
\label{sec:ddr-basics}

Servers predominantly access DRAM over the Double Data Rate (DDR) parallel interface. 
The interface's processor pin requirement is determined by the width of data bus, command/address bus, and configuration pins. 
A DDR4 and DDR5~\cite{JEDEC-DDR5} interface is 288 pins wide. 
While several of those pins are terminated at the motherboard, most of them (160+ for an ECC-enabled DDR4 channel \cite{jaffari:systems}, likely more for DDR5~\cite{rooney2019micron}) are driven to the processor chip. 

The DDR interface's 64 data bits directly connect to the processor and are bit-wise synchronous with the memory controller's clock, %
enabling a worst-case (unloaded) access latency of about 50ns.
Scaling a DDR-based memory system's bandwidth requires either clocking the channels at a higher rate, or attaching more channels to the processors.
The former approach results in signal integrity challenges~\cite{ddr5_design_challenges} and a reduction in supported ranks per channel, limiting rank-level parallelism and  memory capacity.
Accommodating more channels requires more on-chip pins, which cost significant area and power, and complicate placement, routing, and packaging~\cite{zhu:package}. 
Therefore, the pin-count on processor packages has only been doubling about every six years \cite{stanley-marbell:pinned}. 

Thus, reducing the number of cores that contend over a memory channel is difficult without clean-slate technologies, which we discuss in \cref{sec:related}. 
To this end, the emerging CXL interconnect is bound to bridge this gap by leveraging a widely deployed high-bandwidth serial interface, as we discuss next.

\subsection{The High-bandwidth CXL Memory Interconnect}

The Compute Express Link (CXL) %
is a recent interconnect standard, designed to present a unified solution for coherent accelerators, non-coherent devices, and memory expansion devices.
It represents the industry's concerted effort for a standardized interconnect to replace a wide motley collection of proprietary solutions (e.g., OpenCAPI \cite{opencapi-cxl}, Gen-Z \cite{genz-cxl}).
CXL is rapidly garnering industry adoption and is bound to become a dominant interconnect, as PCIe has been for peripheral devices over the past twenty years.

CXL brings load-store semantics and coherent memory access to high-capacity, high-bandwidth memory for processors and accelerators alike. %
It also enables attaching DDR-based memory (``Type-3'' CXL devices) over PCIe to the processor with strict timing constraints. 
In this work, we focus on this capability of CXL.
CXL's underlying PCIe physical layer affords  higher bandwidth per pin at the cost of increased latency.
Therefore, most recent works thus far perceive CXL as a technology enabling an \textit{auxiliary} slower memory tier directly attached to the processor.
In contrast, we argue that despite its associated latency overhead, CXL can play a central role in future memory systems design, \textit{replacing}, rather than simply augmenting, DDR-based memory in server processors. 

\begin{figure}
    \centering

        \includegraphics[width=\columnwidth]{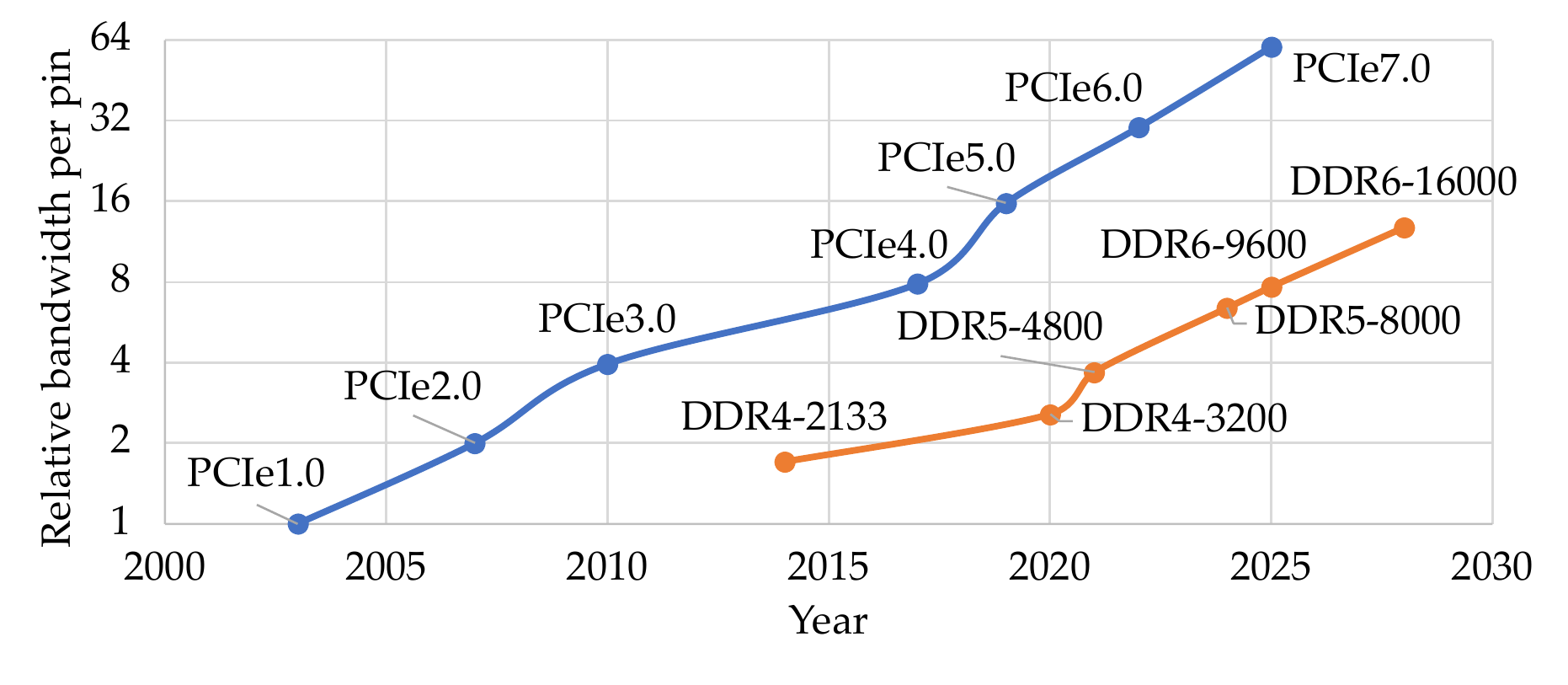}
    
    \vspace{-4mm}
    
    \caption{Bandwidth per processor pin for DDR and CXL (PCIe) interface, norm. to PCIe-1.0. Note that y-axis is in log scale. }
    \label{fig:bandwidth-trends}
    
\end{figure}

\subsection{Scaling the Memory Bandwidth Wall with CXL}
\label{sec:motivation:cxl-bw}

CXL's high bandwidth owes to its underlying PCIe physical layer.
PCIe~\cite{sharma:pci} is a high-speed serial interface featuring multiple independent lanes capable of bi-directional communication using just 4 pins per lane: two for transmitting data, and two for receiving data. 
Data is sent over each lane as a serial bit stream at very high bit rates in an encoded format. 

\cref{fig:bandwidth-trends} illustrates the bandwidth per pin for PCIe and DDR.
The normalized bandwidth per pin is derived by dividing each interface's peak interface bandwidth on  JEDEC's and PCI-SIG's roadmap, respectively, by the processor pins required: 160 for DDR and 4 per lane for PCIe.

The $4\times$ bandwidth gap is where we are today (PCIe5.0 vs. DDR5-4800).
The comparison is conservative, given that PCIe's stated bandwidth is \textit{per direction}, while DDR5-4800 requires about 160 processor pins for a theoretical 38.4GB/s peak of \textit{combined} read and write bandwidth.
With a third of the pins, 12 PCIe5.0 lanes (over which CXL operates) offer 48GB/s per direction---i.e., a theoretical peak of 48GB/s for reads \textit{and} 48GB/s for writes.
Furthermore, \cref{fig:bandwidth-trends}'s roadmaps suggest that the bandwidth gap will grow to $8\times$ by 2025.

\begin{figure*}
    \centering

    \subfloat[Average and p90 memory access latency in a DDR5-4800 channel (38.4GB/s) at varying bandwidth utilization points. p90 grows faster than the average latency.]{
    \includegraphics[width=.7\columnwidth]{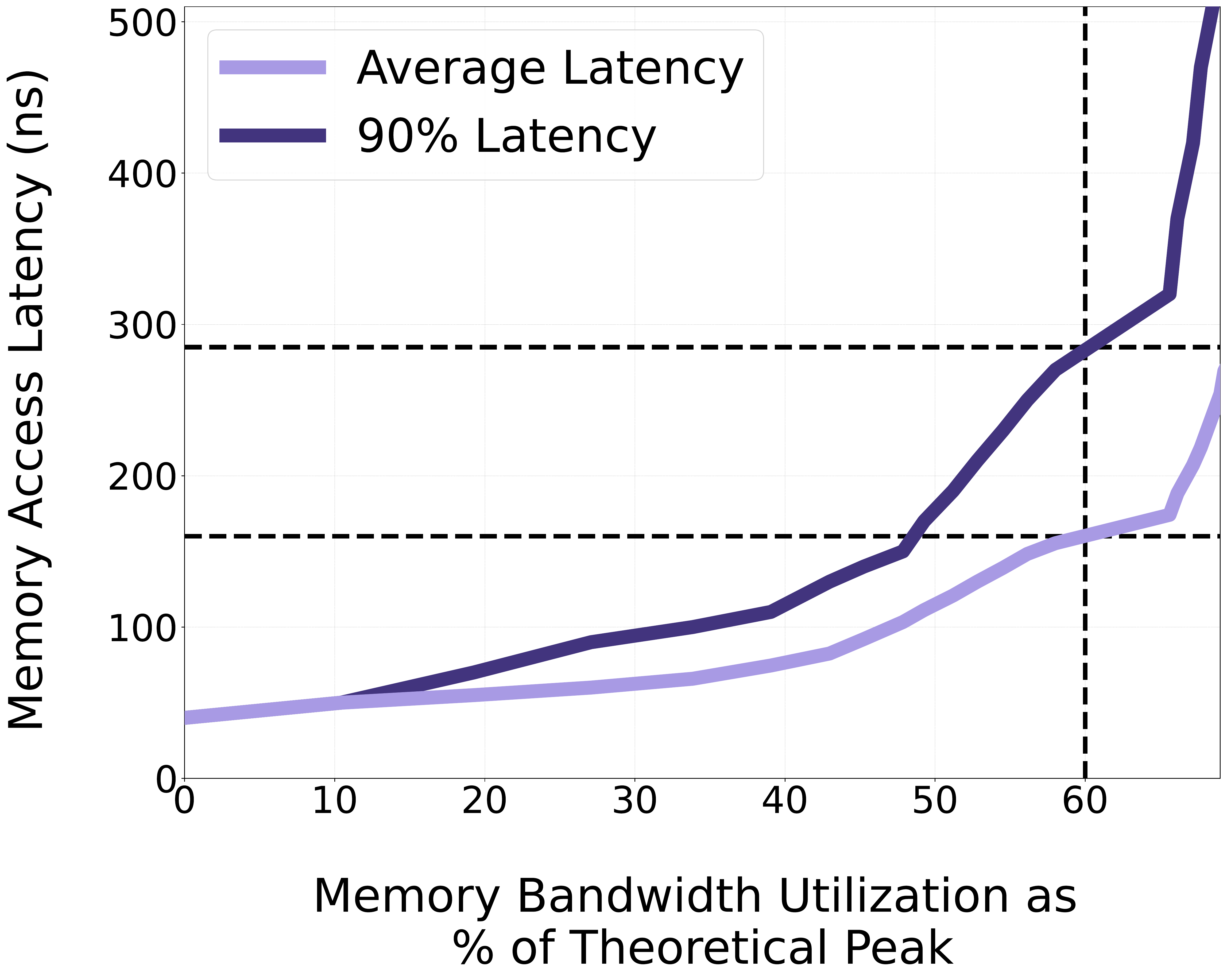}
    \label{fig:latency-bw-motivation}
    }
    \hfill
    \subfloat[Memory latency breakdown (DRAM access time and queuing delay) and memory bandwidth utilization for a range of workloads. Higher utilization increases queuing delay. ]{
     \includegraphics[width=1.23\columnwidth]{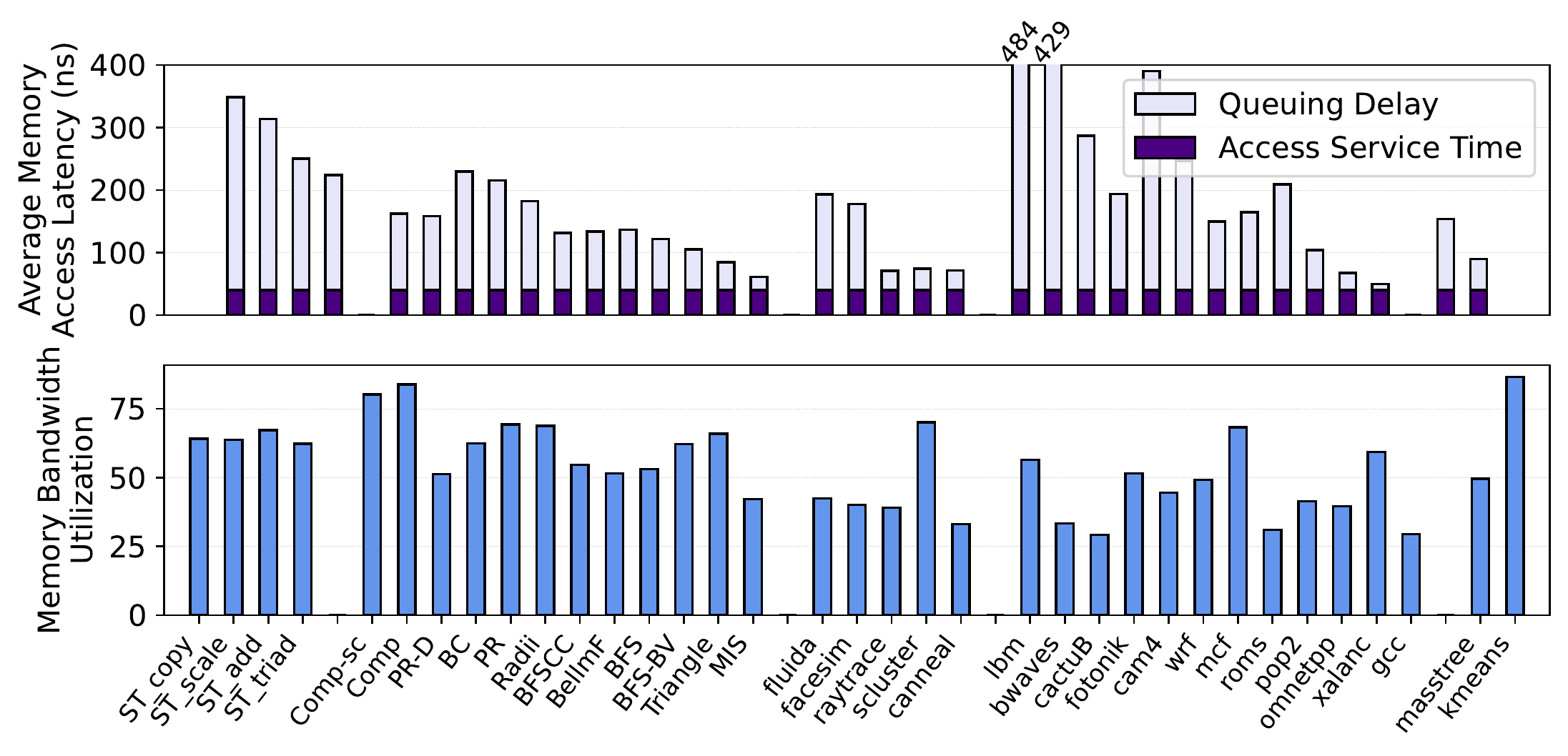}     
         \label{fig:motivation:queuing}
    }
    \caption{Queuing drastically affects memory access time on a loaded system. }
    
\end{figure*}

\subsection{CXL Latency Concerns}
\label{sec:motivation:cxl-lat}

CXL's increased bandwidth comes at the cost of increased latency compared to DDR. 
There is a widespread assumption that this latency cost is significantly higher than DRAM access latency itself.  For instance, recent work on CXL-pooled memory reinforces that expectation, by reporting a latency overhead of 70ns \cite{li:pond}. 
The expectation of such high added latency has reasonably led memory system researchers and designers to predominantly focus on CXL as a technology for enabling a secondary tier of slower memory that augments conventional DDR-attached memory.
However, such a high latency overhead does not represent the minimum attainable latency of the simplest CXL-attached memory and is largely an artifact of more complex functionality, such as multiplexing multiple memory devices, enforcing coherence between the host and memory device, etc.

In this work, we argue that CXL is a perfect candidate to completely \textit{replace} the DDR-attached memory for server processors that handle memory-intensive workloads. 
The CXL 3.0 standard sets an 80ns pin-to-pin load latency target for a CXL-attached memory device \cite[Table 13-2]{cxl-3.0}, which in turn implies that the interface-added latency over DRAM access in upcoming CXL memory devices should be about 30ns. 
Early implementations of the CXL 2.0 standard demonstrated a 25ns latency overhead per direction \cite{plda:breaking}, and in 2021 PLDA announced a commercially available CXL 2.0 controller that only adds 12ns per direction \cite{plda:cxl-latency}. %
Such low latency overheads are attainable with the simplest CXL type-3 devices that are not multiplexed across multiple hosts and do not need to initiate any coherence transactions.
Our key insight is that a memory access latency penalty in the order of 30ns often pales in comparison to queuing delays at the memory controller that are common in server systems, and such queuing delays can be curtailed by CXL's considerable bandwidth boost.

\ifvariance
\section{Pitfalls of Unloaded and Average Latency}
\label{sec:queuing-effect}
\else
\section{Queuing Dictates Effective Memory Access Latency}
\label{sec:motivation:queuing-impact}
\fi

It is evident from current technological trends that systems with CXL-attached memory can enjoy significantly higher bandwidth availability compared to conventional systems with DDR-attached memory.
A key concern hindering broad adoption---and particularly our proposed replacement of DDR interfaces on-chip with CXL---is CXL's increased memory access latency.
However, in any system with a loaded memory subsystem, queuing effects play a significant role in determining effective memory access latency.
\ifvariance
On a loaded system, queuing (i) dominates the effective memory access latency, and (ii) introduces variance in accessing memory,  degrading performance.
We next demonstrate the impact of both effects.

\subsection{Queuing Dictates Effective Memory Access Latency}
\label{sec:motivation:queuing-impact}
\fi

\cref{fig:latency-bw-motivation} shows a DDR5-4800 channel's memory access latency as its load increases.
We model the memory using DRAMSim~\cite{rosenfeld:dramsim2} and control the load with random memory accesses of configurable arrival rate. %
The resulting load-latency curve is shaped by queuing effects at the memory controller.

When the system is unloaded, a hypothetical CXL interface adding 30ns to each memory access would correspond to a seemingly prohibitive 75\% latency overhead compared to the approximated unloaded latency of 40ns. 
However, as the memory load increases, latency rises exponentially, with average latency increasing by $3\times$ and $4\times$ at 50\% and 60\% load, respectively.
p90 tail latency grows even faster, rising by $4.7\times$ and $7.1\times$ at the same load points.
In a loaded system, trading off additional interface latency for considerably higher bandwidth availability can yield significant net latency gain.

To illustrate, consider a baseline DDR-based system operating at 60\% of memory bandwidth utilization, corresponding to 160ns average and 285ns p90 memory access latency.
A CXL-based alternative offering a %
$4\times$ memory bandwidth boost would shrink the system's bandwidth utilization to 15\%, corresponding to 50\% lower average and 68\% {lower} p90 memory access latency compared to baseline, despite the CXL interface's 30ns latency premium. %

\cref{fig:latency-bw-motivation} shows that a system with bandwidth utilization as low as 20\% experiences queuing effects, that are initially reflected on tail latency; beyond 40\% utilization, queuing effects also noticeably affect average latency.
Utilization beyond such level is common, as we show with our simulation of a 12-core processor with 1 DDR5 memory channels over a range of server and desktop applications (methodological details in \cref{sec:method}).
\cref{fig:motivation:queuing} shows that with all processor cores under use, the vast majority of workloads exceed 30\% memory bandwidth utilization, and most exceed 50\% utilization (except several workloads from SPEC and PARSEC benchmarks).

\cref{fig:motivation:queuing} also breaks down the average memory access time seen from the LLC miss register into DRAM service time and queuing delay at the memory controller. 
We observe a trend in high bandwidth consumption leading to long queuing delays, although queuing delay doesn't present itself as a direct function of bandwidth utilization. 
Queuing delay is also affected by application characteristics such as read/write pattern and spatial and temporal distribution of accesses. 
For example, in an access pattern where processor makes majority of memory access requests in a short amount of time, followed by a period of low memory activity, the system would temporarily be in a high bandwidth utilization state when memory requests are made, experiencing contention and high queuing delay, even though the average bandwidth consumption would not be as high. Even in such cases, provisioning more bandwidth would lead to better performance, as it would mitigate contention from the temporary bursts. 
In \cref{fig:motivation:queuing}'s workloads, queuing delay constitutes 72\% of the memory access latency on average, and up to 91\% in the case of \textit{lbm}.

\ifvariance

\subsection{Memory Latency Variance Impacts Performance}
\label{sec:motivation:variance}

In addition to their effect on average memory access latency, spurious queuing effects at the memory controller introduce higher memory access latency fluctuations (i.e., variance). 
Such variance is closely related to the queueing delay stemming from high utilization, as discussed in \cref{sec:motivation:queuing-impact}. 
To demonstrate the impact of memory access latency variance on performance, we conduct a controlled experiment where the average memory access latency is kept constant, but the latency fluctuation around the average grows. 
The baseline is a toy memory system with a 150ns fixed access latency and we evaluate three additional memory systems where memory access latency follows a bimodal distribution with 80\%/20\% probability of being lower/higher than the average. 
We keep average latency constant in all cases ($80\% \times low\_lat + 20\% \times high\_lat = 150ns$) and we evaluate $(low\_lat, high\_lat) $ for $ (100ns, 350ns), (75ns, 450ns), (50ns, 550ns)$, resulting in distributions with increasing standard deviations (stdev) of 100ns, 150ns, and 200ns. Variance is the square of stdev and denotes how spread out the latency is from the average.

\begin{figure} [t]
	\centering
   
      \includegraphics[width=\columnwidth]{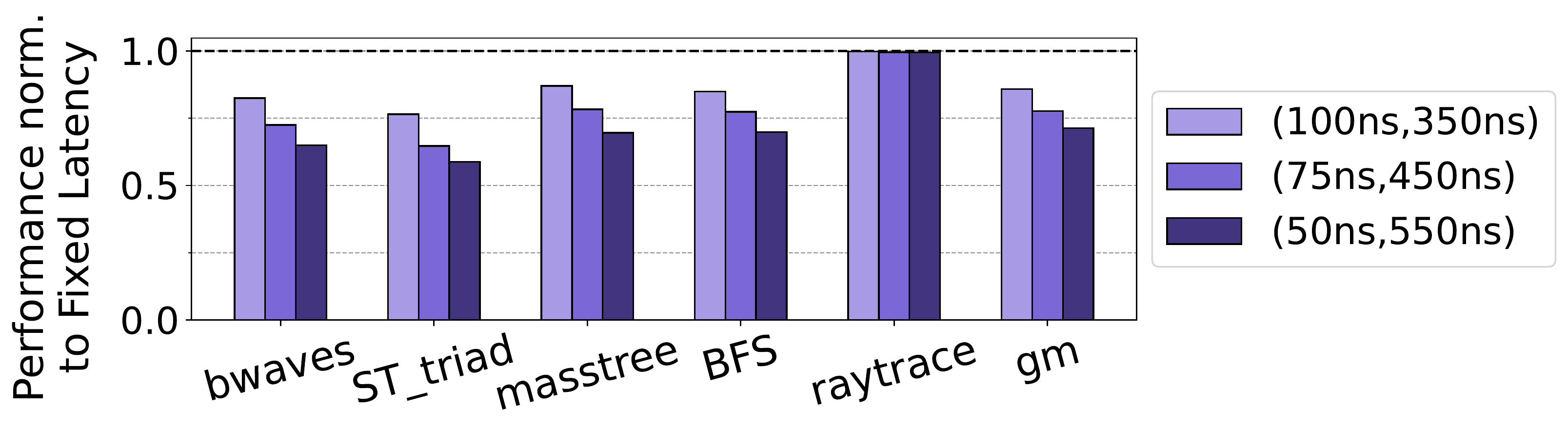}
      \vspace{-4mm}
	\caption{Performance of workloads for synthetic memory access latency following three (X, Y) bimodal distributions with 4:1 X:Y ratio, all with 150ns average latency, normalized to a memory system with fixed 150ns latency. 
   ``gm" refers to geometric mean. Higher latency variance degrades performance. 
 }
	\label{fig:latency-variance-exp}
 \end{figure}  

\cref{fig:latency-variance-exp} shows the relative performance of these memory systems for five workloads of decreasing memory bandwidth intensity. %
As variance increases, the average performance relative to the  fixed-latency baseline noticeably drops to 86\%, 78\%, and 71\%. %
This experiment highlights that solely relying on typical average metrics like Average Memory Access Time (AMAT) to assess a memory system's performance is an incomplete method of evaluating a memory system's performance. 
In addition to average values, the variance of memory access latency is a major performance determinant and therefore an important quality criterion for a memory system.

\fi

\vspace{0.05in} 
\begin{tcolorbox}[colframe=blue!75!black,colback=white,boxrule=1pt,left=5pt,right=5pt,top=3pt,bottom=3pt]
\noindent\textbf{Summary:} 
The significant memory bandwidth boost attainable with CXL-attached memory more than compensates for the interface's higher latency in loaded systems, where queuing  dictates the memory system's effective access latency.
By decreasing queuing, a CXL-based memory system reduces average memory access time and variance, both of which improve performance.
\end{tcolorbox}

\begin{table*}
\begin{footnotesize}
\def\arraystretch{1.1}
    \begin{minipage}{.22\textwidth}
        \centering
            \caption{Area of processor components at TSMC 7nm (rel. to 1MB of L3 cache).}
\label{table:relsize}
\begin{tabular}{ | l | c | }
\hline
L3 cache (1MB) & 1\\
\hline
Zen 3 Core  &	\multirow{2}{*}{6.5}	\\
(incl. 512 KB L2) & \\
 \hline
x8 PCIe (PHY + ctrl)	& \multirow{1}{*}{5.9} \\
\hline
DDR channel (PHY + ctrl) &	\multirow{1}{*}{10.8}\\
\hline
\end{tabular}
    \end{minipage}%
    \hspace{6mm}
    \begin{minipage}{.75\textwidth}
        \centering
        \caption{DDR-based versus alternative \TheName server configurations.}
\label{table:server-configs}
\begin{tabular}{|c|c|c|c|c|c|ll|}
\hline
\multirow{2}{*}{\textbf{Server design}} & \textbf{Core} & \textbf{LLC} & \textbf{Memory} & \textbf{Relative} & \textbf{Relative} & \multicolumn{2}{c|}{\multirow{2}{*}{\textbf{Comment}}}                                          \\ 
 & \textbf{count} & \textbf{per core} & \textbf{interfaces} & \textbf{mem. BW} & \textbf{area} & \multicolumn{2}{c|}{}\\
\hline
DDR-based                       & \multirow{6}{*}{144} & \multirow{3}{*}{2 MB}    & 12 DDR                      & 1$\times$                       & 1                  & \multicolumn{2}{l|}{baseline}                                                  \\ \cline{1-1} \cline{4-8} 
\TheName-$5\times$              &                     &                       & 60 x8 CXL                  & 5$\times$                    & 1.17                & \multicolumn{2}{l|}{iso-pin}                                                   \\ \cline{1-1} \cline{4-8} 
\TheName-$2\times$              &                     &                       & 24 x8 CXL                  & 2$\times$                        & \multicolumn{1}{c|}{\multirow{4}{*}{1.01}}               & \multicolumn{1}{l|}{\multirow{4}{*}{iso-area}} & iso-LLC                       \\ \cline{1-1} \cline{3-5} \cline{8-8} 
\TheName-$4\times$              &                     & \multirow{3}{*}{1 MB}   & 48 x8 CXL                  & 4$\times$                        &                & \multicolumn{1}{l|}{}                          & \multicolumn{1}{c|}{balanced} \\ \cline{1-1} \cline{4-5} \cline{8-8} 
\multirow{2}{*}{\TheName-asym}  &                     &                         & 48 x8 CXL-asym         & asym. R/W                        &                & \multicolumn{1}{l|}{}                          & \multicolumn{1}{c|}{\multirow{2}{*}{max BW}} \\ %
                                &                      &                        & \multicolumn{2}{c|}{(see \cref{sec:coaxial-opt})}         &                  & \multicolumn{1}{l|}{}   & \\
\cline{1-8}

\end{tabular} 
    \end{minipage}
\end{footnotesize}
\end{table*}

\section{The \TheName Server Architecture}
\label{sec:design}

\begin{figure} %
	\centering
   
    \captionsetup[subfloat]{captionskip=-12pt}   	
	\subfloat[Baseline DDR-based server. %
        ]{
            \includegraphics[width=.85\columnwidth]{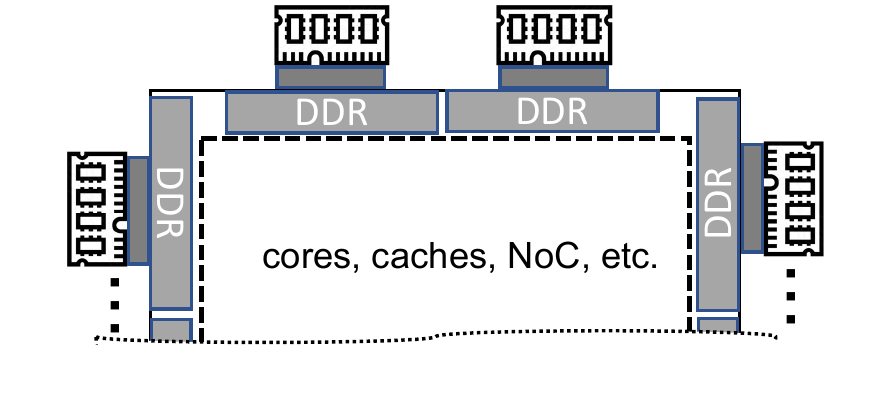}
		\label{fig:system-architecture:ddr}} %
            \vspace{7mm}
        \captionsetup[subfloat]{captionskip=10pt, width=\linewidth}         
	\subfloat[\TheName replaces each DDR channel with several CXL channels. Each CXL channel connects to a type-3 device with one DDR memory channel.] {
            \includegraphics[width=.9\columnwidth, trim=2mm 8mm 2mm 12mm]{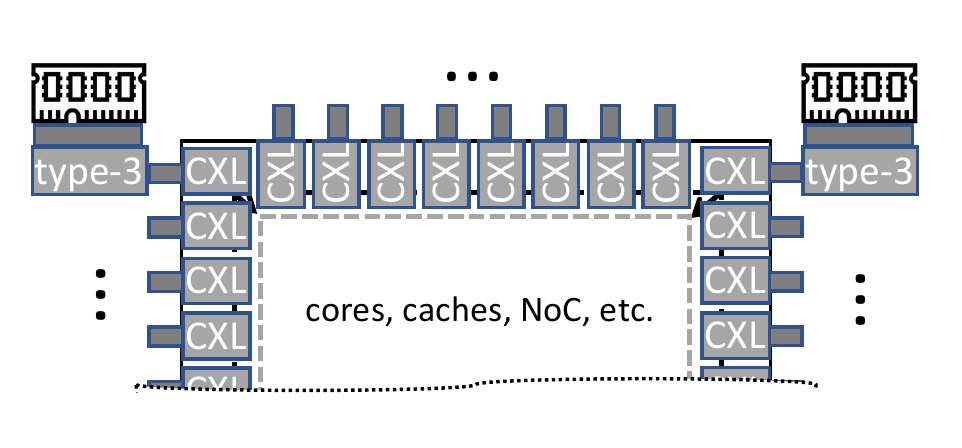}
		\label{fig:system-architecture:cxl}}
        \caption{Overview of the baseline and \TheName systems. }
	\label{fig:system-architecture}
 \end{figure}

We leverage CXL's per-pin bandwidth advantage to replace \textit{all} of the DDR interfaces with PCIe-based CXL interfaces in our proposed \TheName server. 
\cref{fig:system-architecture:cxl} depicts our architecture where each on-chip DDR5 channel is replaced by several CXL channels, providing 2--4$\times$ higher aggregate memory bandwidth to the processor. 
\cref{fig:system-architecture:ddr} shows the baseline DDR-based server design for comparison. 
Each CXL channel is attached to a ``Type-3`` CXL device, which features a memory controller that manages a regular DDR5 channel that connects to DRAM. %
The processor implements the {\tt CXL.mem} protocol of the CXL standard, which orchestrates data consistency and memory semantics management. The implementation of the caches and cores remains unchanged, as the memory controller still supplies 64B cache lines. 

\subsection{Processor Pin Considerations}

A DDR5-4800 channel features a peak \textit{uni-directional} bandwidth of 38.4GB/s and requires more than 160 processor pins to account for data and ECC bits, command/address bus, data strobes, clock, feature modes, etc., as described in \cref{sec:ddr-basics}. A full 16-lane PCIe connection delivers 64GB/s of \textit{bi-directional} bandwidth. Moreover, PCIe is modular, and higher-bandwidth channels can be constructed by grouping independent lanes together. Each lane requires just four processor pins: two each for transmitting and receiving data. 

The PCIe standard currently only allows groupings of 1, 2, 4, 8, 12, 16, or 32 lanes.  
To match DDR5's bandwidth of 38.4GB/s, we opt for an x8 configuration, which requires 32 pins for a peak bandwidth of 32GB/s, $5\times$ fewer than the 160 pins required for the DDR5 channel. 
As PCIe can sustain 32GB/s bandwidth in each direction, the peak aggregate bandwidth of 8 lanes is 64GB/s, much higher than DDR5's 38.4GB/s. 
Considering a typical 2:1 Read:Write ratio, only 25.6GB/s of a DDR5 channel's bandwidth would be used in the DRAM-to-CPU direction, and about 13GB/s in the opposite direction.
Furthermore, peak sustainable bandwidth for DDR controllers typically achieve around 70\% to 90\% of the theoretical peak.
Thus, even after factoring in PCIe and CXL's header overheads which reduce the practically attainable bandwidth \cite{sharma:compute} to 26GB/s in the DRAM-to-CPU direction and 13GB/s in the other direction, the x8 configuration supports a full DDR5 channel without becoming a choke point.

\subsection{Silicon Area Considerations}

When it comes to processor pin requirements, \TheName allows replacement of each DDR channel (i.e., PHY and memory controller) with five x8 PCIe PHY and controllers, for a $5\times$ memory bandwidth boost.
However, the relative pin requirements of DDR and PCIe are not directly reflected in their relative on-chip silicon area requirements.
Lacking publicly available information, we derive the relative size of DDR and PCIe PHYs and controllers from AMD Rome and Intel Golden Cove die shots \cite{intel-die-shot, amd-die-shot}. 

\cref{table:relsize} shows the relative silicon die area different key components of the processor account for. 
Assuming linear scaling of PCIe area with the number of lanes, as appears to be the case from the die shots, an x8 PCIe controller accounts for 54\% of a DDR controller's area. 
Hence, replacing each DDR controller with four x8 PCIe controllers requires 2.19$\times$ more silicon area than what is allocated to DDR.
However, DDR controllers account for a small fraction of the total CPU die. %

Leveraging \cref{table:relsize}'s information, we now consider a number of alternative \TheName server designs, shown in \cref{table:server-configs}.
We focus on high-core-count servers optimized for throughput, such as the  upcoming %
AMD EPYC Bergamo (128 cores)~\cite{wccf_bergamo}, and Intel Granite Rapids (128 cores) and Sierra Forest (144 cores) \cite{wccf_intel_sierra}. 
All of them feature 12 DDR5 channels, resulting in a core-to-memory-controller (core:MC) ratio of 10.7:1 to 12:1. 
A common design choice to accommodate such high core counts is a reduced LLC capacity; e.g., moving from the 96-core Genoa~\cite{amd-genoa} to the 128-core Bergamo, AMD halves the LLC per core to 2MB.
We thus consider a 144-core baseline server processor with 12 DDR5 channels and 2MB of LLC per core (\cref{table:server-configs}, first row).

With pin count as its only limitation, \textit{\TheName-$5\times$} replaces each DDR channel with 5 x8 CXL interfaces, for a $5\times$ bandwidth increase. 
Unfortunately, that results in a 17\% increase in die area to accommodate all the PCIe PHYs and controllers.
Hence, we also consider two iso-area alternatives.
\textit{\TheName-$2\times$} leverages CXL to double memory bandwidth without any microarchitectural changes.
\textit{\TheName-$4\times$} quadruples the available memory bandwidth compared to the baseline CPU by halving the LLC from 288MB to 144MB.

\subsection{\TheName Asymmetric Interface Optimization}
\label{sec:coaxial-opt}

A key difference between CXL and DDR is that the former provisions dedicated pins and wires for each data movement direction (RX and TX).
The PCIe standard defines a one-to-one match of TX and RX pins: e.g., an x8 PCIe configuration implies 8 TX and 8 RX lanes.
We observe that while uniform bandwidth provisioning in each direction is reasonable for a peripheral device like a NIC, it is not the case for memory traffic. 
Because (i) most workloads read more data than they write and (ii) every cache block that is written must typically be read first, R:W ratios are usually in the 3:1 to 2:1 range rather than 1:1. 
Thus, in the current 1:1 design, read bandwidth becomes the bottleneck and write bandwidth is underutilized.
Given this observation and that serial interfaces do not fundamentally require 1:1 RX:TX bandwidth provisioning~\cite{wang:alloy}, we consider a \TheName design with asymmetric RX/TX lane provisioning to better match memory traffic characteristics. 
While the PCIe standard currently disallows doing so, we investigate the potential performance benefits of revisiting that  statutory restriction.
We call such a channel \textit{CXL-asym}.

We consider a system leveraging such CXL-asym channels to compose an additional \textbf{\TheName-asym} configuration.
An x8 CXL channel consists of 32 pins, 16 each way. Without the current 1:1 PCIe restriction, 
CXL-asym repurposes the same pin count to use 20 RX pins and 12 TX pins, resulting in 40GB/s RX and 24GB/s TX of raw bandwidth. Accounting for PCIe and CXL's header overheads, the realized bandwidth is approximately 32GB/s for reads (compared to 26GB/s in x8 CXL channel) and 10GB/s for writes \cite{sharma:compute}. 
To utilize the additional read bandwidth, we provision two DDR controllers per CXL-asym channel on the type-3 device.
Therefore, the number of CXL channels on the processor (as well as their area overhead) remains unchanged.
While the 32GB/s read bandwidth of CXL-asym is insufficient to support two DDR channels at their combined read bandwidth of about 52GB/s (assuming a 2:1 R:W ratio), queuing delays at the DDR controller typically become significant at a much lower utilization point, as shown in \cref{fig:latency-bw-motivation}.
Therefore, \TheName-asym still provides sufficient bandwidth to eliminate contention at queues by lowering the overall bandwidth utilization, while providing higher aggregate bandwidth.

\subsection{Additional Benefits of \TheName}

Our analysis focuses on the performance impact of a CXL-based memory system.
While a memory capacity and cost analysis is beyond the scope of this paper, \TheName can have additional positive effects on those fronts that are noteworthy. 
Servers provisioned for maximum memory capacity deploy two high-density DIMMs per DDR channel. The implications are two-fold. First, two-DIMMs-per-channel (2DPC) configurations increase capacity over 1DPC at the cost of $\sim$15\% memory bandwidth.
Second, DIMM cost grows superlinearly with density; for example, 128GB/256GB DIMMs cost 5$\times$/20$\times$ more than 64GB DIMMs.
By enabling more DDR channels, \TheName allows the same or higher DRAM capacity with 1DPC and lower-density DIMMs.

\section{Evaluation Methodology}
\label{sec:method}

\begin{table}[t]
\begin{center}
\begin{footnotesize}
\def\arraystretch{1.2}  
\caption{System parameters used for simulation on \simulator.}

\begin{tabular}{|p{1cm} | p{1.9cm}  p{2.1cm}  p{1.8cm}|}
\hline
 & \multicolumn{1}{c|}{DDR baseline} & \multicolumn{2}{c|}{CoaXiaL-*}  \\ \hline
CPU & \multicolumn{3}{l|}{12 OoO cores, 2GHz, 4-wide, 256-entry ROB} \\ \hline
L1 & \multicolumn{3}{l|}{32KB L1-I \& L1-D, 8-way, 64B blocks, 4-cycle access} \\ \hline
L2 & \multicolumn{3}{l|}{512 KB, 8-way, 12-cycle access} \\ \hline
\multirow{2}{*}{LLC} & \multicolumn{3}{l|}{shared \& non-inclusive, 16-way, 46-cycle access} \\ \cline{2-4} 
 & \multicolumn{1}{l|}{2 MB/core} & \multicolumn{2}{l|}{1--2 MB/core (see \cref{table:server-configs})} \\ \hline
\multirow{3}{*}{Memory} & \multicolumn{3}{l|}{\begin{tabular}[c]{@{}l@{}}DDR5-4800 \cite{micron_ddr5}, 128 GB per channel, 2 sub-channels\\ per channel, 1 rank per sub-channel, 32 banks per rank\end{tabular}} \\ \cline{2-4} 
 & \multicolumn{1}{l|}{\begin{tabular}[c]{@{}l@{}}1 channel\end{tabular}} & \multicolumn{2}{l|}{\begin{tabular}[c]{@{}l@{}}2--4 CXL-attached channels (see \cref{table:server-configs})\\8 channels for \TheName-asym (see \cref{sec:coaxial-opt})\end{tabular}} \\ \hline
\end{tabular}
\label{table:simSetup}
\end{footnotesize}
\end{center}
\end{table}

\smallskip
\noindent\textbf{System configurations.} 
We compare our \TheName server design, which replaces the processor's DDR channels with CXL channels, to a typical DDR-based server processor. %
\begin{itemize}[noitemsep,topsep=0pt,leftmargin=*]
    \item \textit{DDR-based baseline}. We simulate 12 cores and one DDR5-4800 memory channel as a scaled-down version of \cref{table:server-configs}'s 144-core CPU. 
    \item \textit{\TheName servers}. We evaluate several servers that replace the on-chip DDR interfaces with CXL: \TheName-2$\times$, \TheName-4$\times$, and \TheName-asym  (\cref{table:server-configs}). %
\end{itemize}
We simulate the above system configurations using \simulator~\cite{champsim} coupled with DRAMsim3 \cite{li:dramsim3}. %
\cref{table:simSetup} summarizes the configuration parameters used.

\smallskip
\noindent\textbf{CXL performance modeling.}
For \TheName, we model CXL controllers and PCIe bus on both the processor and the type-3 device.
Each CXL controller comprises a CXL port that incurs a fixed delay of 12ns accounting for flit-packing, encoding-decoding, packet processing, etc. \cite{plda:cxl-latency}.  
The PCIe bus incurs traversal latency due to the limited channel bandwidth and bus width.
For an x8 channel, the peak 32GB/s bandwidth results in 26/13 GB/s RX/TX goodput when header overheads are factored in, and 32/10 GB/s RX/TX in the case of CXL-asym channels. 
The corresponding link traversal latency is 2.5/ 5.5 ns RX/TX for an x8 channel and 2/ 9 ns RX/TX for CXL-asym.
Additionally, the CXL controller maintains message queues to buffer requests.
Therefore, in addition to minimum latency overhead of about 30ns (or more, in our sensitivity analysis), queuing effects at the CXL controller are also captured and reflected in the performance.

\smallskip
\noindent\textbf{Workloads.} We evaluate 35 workloads from various benchmark suites. We deploy the same workload instance on all cores and simulate 200 million instructions per core after fast-forwarding each application to a region of interest.
\begin{itemize}[noitemsep,topsep=0pt,leftmargin=*]
    \item \textit{Graph analytics:} We use 12 workloads from the LIGRA benchmark suite \cite{shun:ligra}. %
    \item \textit{STREAM:} We run the four kernels (\textit{copy, scale, add, triad}) from the STREAM benchmark \cite{mccalpin:memory} to represent bandwidth-intensive matrix operations in which ML workloads spend  a significant portion of their execution time.
    \item \textit{SPEC \& PARSEC:} We evaluate 13 workloads from the SPEC-speed 2017~\cite{SPEC2017} benchmark suite in \textit{ref} mode, %
    as well as five  PARSEC workloads \cite{bienia:parsec}.
    \item We evaluate \textit{masstree} \cite{mao:cache} and \textit{kmeans} \cite{lloyd:least} to represent key value store and data analytics workloads, respectively. 
\end{itemize}

\noindent \cref{table:apps} summarizes all our evaluated workloads, along with their IPC and MPKI as measured on the DDR-based baseline.

\begin{table}%
\begin{center}
\begin{footnotesize}
\def\arraystretch{1.1}  
\caption{Workload Summary.}
\label{table:apps}
\begin{tabular}{|lcc|lcc}
\hline
\multicolumn{1}{|l|}{\textbf{Application}} & \multicolumn{1}{l|}{\textbf{IPC}} & \multicolumn{1}{l||}{\textbf{\begin{tabular}[c]{@{}l@{}}LLC\\ MPKI\end{tabular}}} & \multicolumn{1}{l|}{\textbf{Application}} & \multicolumn{1}{l|}{\textbf{IPC}} & \multicolumn{1}{l|}{\textbf{\begin{tabular}[c]{@{}l@{}}LLC\\ MPKI\end{tabular}}} \\ \hline
\multicolumn{3}{|c||}{\textbf{Ligra}} & \multicolumn{3}{c|}{\textbf{SPEC}} \\ \hline
\multicolumn{1}{|l|}{PageRank} & \multicolumn{1}{c|}{0.36} & \multicolumn{1}{c||}{40} & \multicolumn{1}{l|}{lbm} & \multicolumn{1}{c|}{0.14} & \multicolumn{1}{c|}{64} \\ \hline
\multicolumn{1}{|l|}{\begin{tabular}[c]{@{}l@{}}PageRank\\ Delta\end{tabular}} & \multicolumn{1}{c|}{0.31} & \multicolumn{1}{c||}{27} & \multicolumn{1}{l|}{bwaves} & \multicolumn{1}{c|}{0.33} & \multicolumn{1}{c|}{14} \\ \hline
\multicolumn{1}{|l|}{\begin{tabular}[c]{@{}l@{}}Components\\ -shortcut\end{tabular}} & \multicolumn{1}{c|}{0.34} & \multicolumn{1}{c||}{48} & \multicolumn{1}{l|}{cactusBSSN} & \multicolumn{1}{c|}{0.68} & \multicolumn{1}{c|}{8} \\ \hline
\multicolumn{1}{|l|}{Components} & \multicolumn{1}{c|}{0.36} & \multicolumn{1}{c||}{48} & \multicolumn{1}{l|}{fotonik3d} & \multicolumn{1}{c|}{0.33} & \multicolumn{1}{c|}{22} \\ \hline
\multicolumn{1}{|l|}{BC} & \multicolumn{1}{c|}{0.33} & \multicolumn{1}{c||}{34} & \multicolumn{1}{l|}{cam4} & \multicolumn{1}{c|}{0.87} & \multicolumn{1}{c|}{6} \\ \hline
\multicolumn{1}{|l|}{Radii} & \multicolumn{1}{c|}{0.41} & \multicolumn{1}{c||}{33} & \multicolumn{1}{l|}{wrf} & \multicolumn{1}{c|}{0.61} & \multicolumn{1}{c|}{11} \\ \hline
\multicolumn{1}{|l|}{BFSCC} & \multicolumn{1}{c|}{0.68} & \multicolumn{1}{c||}{17} & \multicolumn{1}{l|}{mcf} & \multicolumn{1}{c|}{0.793} & \multicolumn{1}{c|}{13} \\ \hline
\multicolumn{1}{|l|}{BFS} & \multicolumn{1}{c|}{0.69} & \multicolumn{1}{c||}{15} & \multicolumn{1}{l|}{roms} & \multicolumn{1}{c|}{0.783} & \multicolumn{1}{c|}{6} \\ \hline
\multicolumn{1}{|l|}{BFS-Bitvector} & \multicolumn{1}{c|}{0.84} & \multicolumn{1}{c||}{15} & \multicolumn{1}{l|}{pop2} & \multicolumn{1}{c|}{1.55} & \multicolumn{1}{c|}{3} \\ \hline
\multicolumn{1}{|l|}{BellmanFord} & \multicolumn{1}{c|}{0.86} & \multicolumn{1}{c||}{9} & \multicolumn{1}{l|}{omnetpp} & \multicolumn{1}{c|}{0.51} & \multicolumn{1}{c|}{10} \\ \hline
\multicolumn{1}{|l|}{Triangle} & \multicolumn{1}{c|}{0.65} & \multicolumn{1}{c||}{21} & \multicolumn{1}{l|}{xalancbmk} & \multicolumn{1}{c|}{0.55} & \multicolumn{1}{c|}{12} \\ \hline
\multicolumn{1}{|l|}{MIS} & \multicolumn{1}{c|}{1.37} & \multicolumn{1}{c||}{8} & \multicolumn{1}{l|}{gcc} & \multicolumn{1}{c|}{0.31} & \multicolumn{1}{c|}{19} \\ \hline
\multicolumn{3}{|c||}{\textbf{STREAM}} & \multicolumn{3}{c|}{\textbf{PARSEC}} \\ \hline
\multicolumn{1}{|l|}{Stream-copy} & \multicolumn{1}{c|}{0.17} & \multicolumn{1}{c||}{58} & \multicolumn{1}{l|}{fluidanimate} & \multicolumn{1}{c|}{0.78} & \multicolumn{1}{c|}{7} \\ \hline
\multicolumn{1}{|l|}{Stream-scale} & \multicolumn{1}{c|}{0.21} & \multicolumn{1}{c||}{48} & \multicolumn{1}{l|}{facesim} & \multicolumn{1}{c|}{0.74} & \multicolumn{1}{c|}{6} \\ \hline
\multicolumn{1}{|l|}{Stream-add} & \multicolumn{1}{c|}{0.16} & \multicolumn{1}{c||}{69} & \multicolumn{1}{l|}{raytrace} & \multicolumn{1}{c|}{1.17} & \multicolumn{1}{c|}{5} \\ \hline
\multicolumn{1}{|l|}{Stream-triad} & \multicolumn{1}{c|}{0.18} & \multicolumn{1}{c||}{59} & \multicolumn{1}{l|}{streamcluster} & \multicolumn{1}{c|}{0.99} & \multicolumn{1}{c|}{14} \\ \hline
\multicolumn{3}{|c||}{\textbf{KVS \& Data analytics}} & \multicolumn{1}{l|}{canneal} & \multicolumn{1}{c|}{0.66} & \multicolumn{1}{c|}{7} \\ \hline
\multicolumn{1}{|l|}{Masstree} & \multicolumn{1}{c|}{0.37} & \multicolumn{1}{c||}{21} &  & \multicolumn{1}{l}{} & \multicolumn{1}{l}{} \\ \cline{1-3}
\multicolumn{1}{|l|}{Kmeans} & \multicolumn{1}{c|}{0.50} & \multicolumn{1}{c||}{36} &  & \multicolumn{1}{l}{} & \multicolumn{1}{l}{} \\ \cline{1-3}
\end{tabular}

\end{footnotesize}
\end{center}
\vspace{-4mm}
\end{table}

\section{Evaluation Results}
\label{sec:eval}

We first compare our main \TheName design, \TheName-$4\times$, with the DDR-based baseline  by analyzing the impact of reduced bandwidth utilization and queuing delays on performance in \cref{sec:eval:all_apps}\footnote{Note that, throughout \cref{sec:eval}'s evaluation, reference to ``\TheName'' without a following qualifier implies the \TheName-$4\times$ configuration.}.
\cref{sec:eval:cdfs} highlights the effect of memory access pattern and distribution on performance.
\cref{sec:eval:bw-boost-sensitivity} presents the performance of alternative \TheName designs, \TheName-$2\times$ and \TheName-asym, and \cref{sec:eval:lat-sensitivity} demonstrates the impact of a more conservative 50ns CXL latency penalty.
\cref{sec:eval:unloaded} evaluates \TheName at different server utilization points, %
and \cref{sec:eval:power} analyzes \TheName's power implications.

\begin{figure*}
    \centering

    \includegraphics[width=\linewidth]{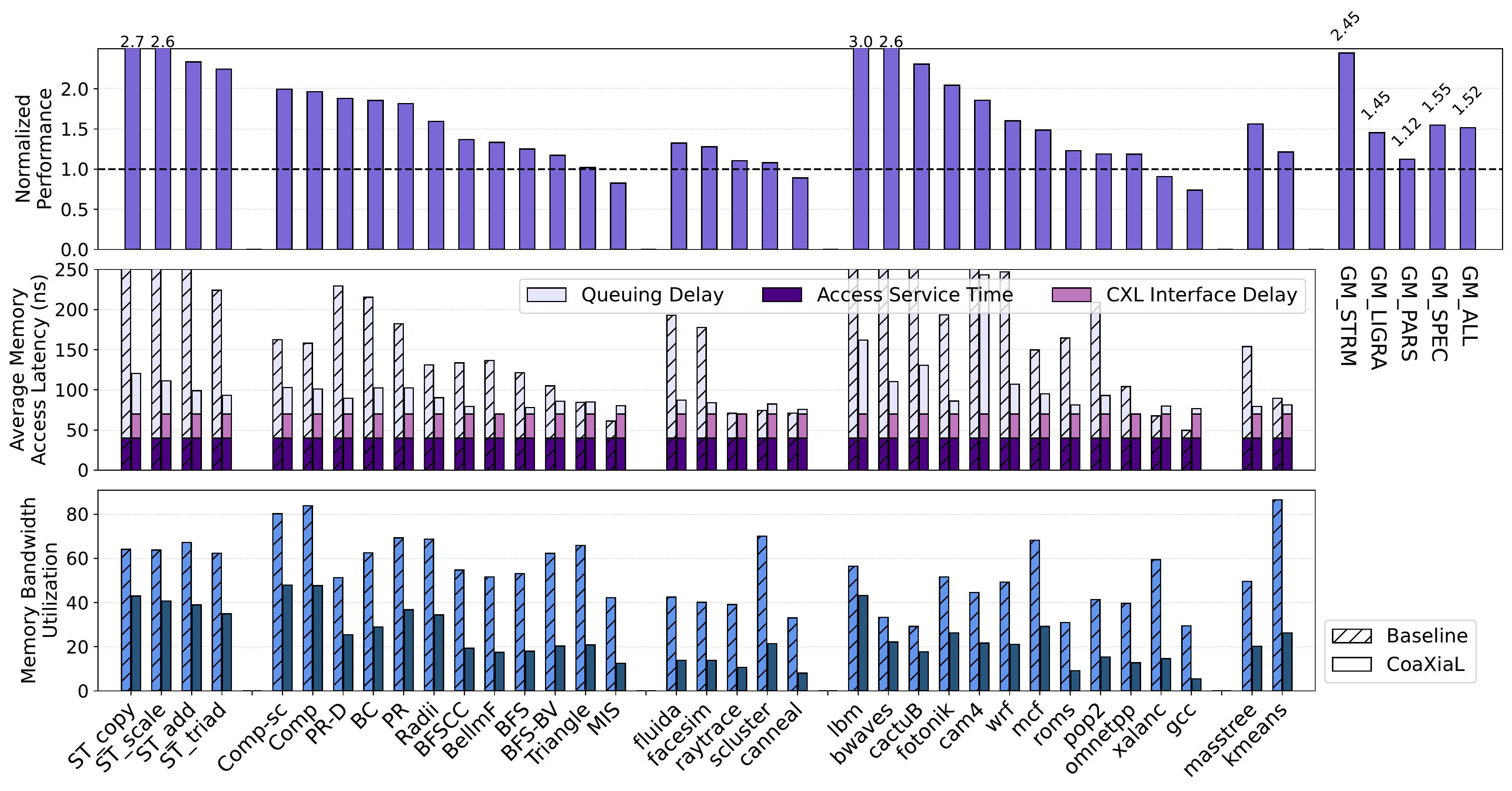}
    
    \vspace{-2mm}
    
    \caption{Normalized performance of \TheName over DDR-based baseline (top), memory access latency breakdown (middle), and memory bandwidth utilization (bottom). Workloads are grouped into their benchmark suite. ``gm'' refers to geometric mean. \TheName offers $1.52\times$ average speedup due to $4\times$ higher bandwidth, lowering utilization and mitigating queuing effects.  }
    \label{fig:main-eval}
    
\end{figure*}

\subsection{From Queuing Reduction to Performance Gains}
\label{sec:eval:all_apps}

\cref{fig:main-eval} (top) shows the performance of \TheName-$4\times$ relative to the baseline DDR-based system.
Most workloads exhibit significant speedup, up to $3\times$ for \textit{lbm} and $1.52\times$ on average. 10 of the 35 workloads experience more than $2\times$ speedup. 
Four workloads lose performance, with gcc most significantly impacted at 26\% IPC loss. 
Workloads most likely to suffer a performance loss are those with  low to moderate memory traffic and heavy dependencies among memory accesses.

\cref{fig:main-eval} (bottom) shows memory bandwidth utilization for the DDR-based baseline and \TheName-$4\times$, which provides $4\times$ higher bandwidth than the baseline.
\TheName distributes memory requests over more channels which reduces the bandwidth utilization of the system, in turn reducing contention for the memory bus.
The lower bandwidth utilization and contention drastically reduces the queuing delay in \TheName for memory-intensive workloads.
\cref{fig:main-eval} (middle) demonstrates this reduction with a breakdown of the average memory access latency (as measured from the LLC miss register) into the DRAM service time, queuing delay, and CXL interface delay (only applicable to \TheName).

In many cases, \TheName enables the workload to drive significantly more aggregate bandwidth from the system.
For instance, \textit{stream-copy} is bottlenecked by the  baseline system's constrained bandwidth, resulting in average queuing delay exceeding 300ns that largely dictates the overall access latency (the total height of the stacked bars).
\TheName reduces queuing delay to just 55ns for this workload, more than compensating for the 30ns CXL interface latency overhead.
The overall average access latency for \textit{stream-copy} reduces from 348ns in baseline to just 120ns, enabling \TheName  to drive memory requests at a $2.9\times$ higher rate versus the baseline, thus achieving commensurate speedup.

Despite provisioning $4\times$ more bandwidth, \TheName reduces average bandwidth utilization from 54\%  to 34\% for workloads that have more than $2\times$ performance improvement, highlighting the extra bandwidth is indeed utilized by these workloads.
For most of the other workloads,  \TheName's average memory access latency is much lower than the baseline's,
despite the CXL interface's latency overhead.

On average, workloads experience 144ns in queuing delay on top of $\sim$40ns DRAM service time.
By slashing queuing delay to just 31ns on average, \TheName reduces average  memory access latency, thereby boosting performance.
Overall, \cref{fig:main-eval}'s results confirm our key insight (see \cref{sec:motivation:queuing-impact}): queuing delays largely dictate the average memory access latency.

\newcounter{takeaway}
\setcounter{takeaway}{0}
\vspace{0.05in} 
\stepcounter{takeaway}
\begin{tcolorbox}[colframe=blue!75!black,colback=white,boxrule=1pt,left=5pt,right=5pt,top=3pt,bottom=3pt]
\textbf{Takeaway \#\thetakeaway{}:} \TheName drastically reduces queuing delays, resulting in lower effective memory access latency for bandwidth-hungry workloads.
\end{tcolorbox}

\subsection{Beyond Average Bandwidth Utilization and Access Latency}
\label{sec:eval:cdfs}

\begin{figure}
    \centering
    \vspace{-3mm}

    \subfloat[Average memory access latency per workload group, and stdev shown as error bars.]{
    \includegraphics[width=.45\columnwidth]{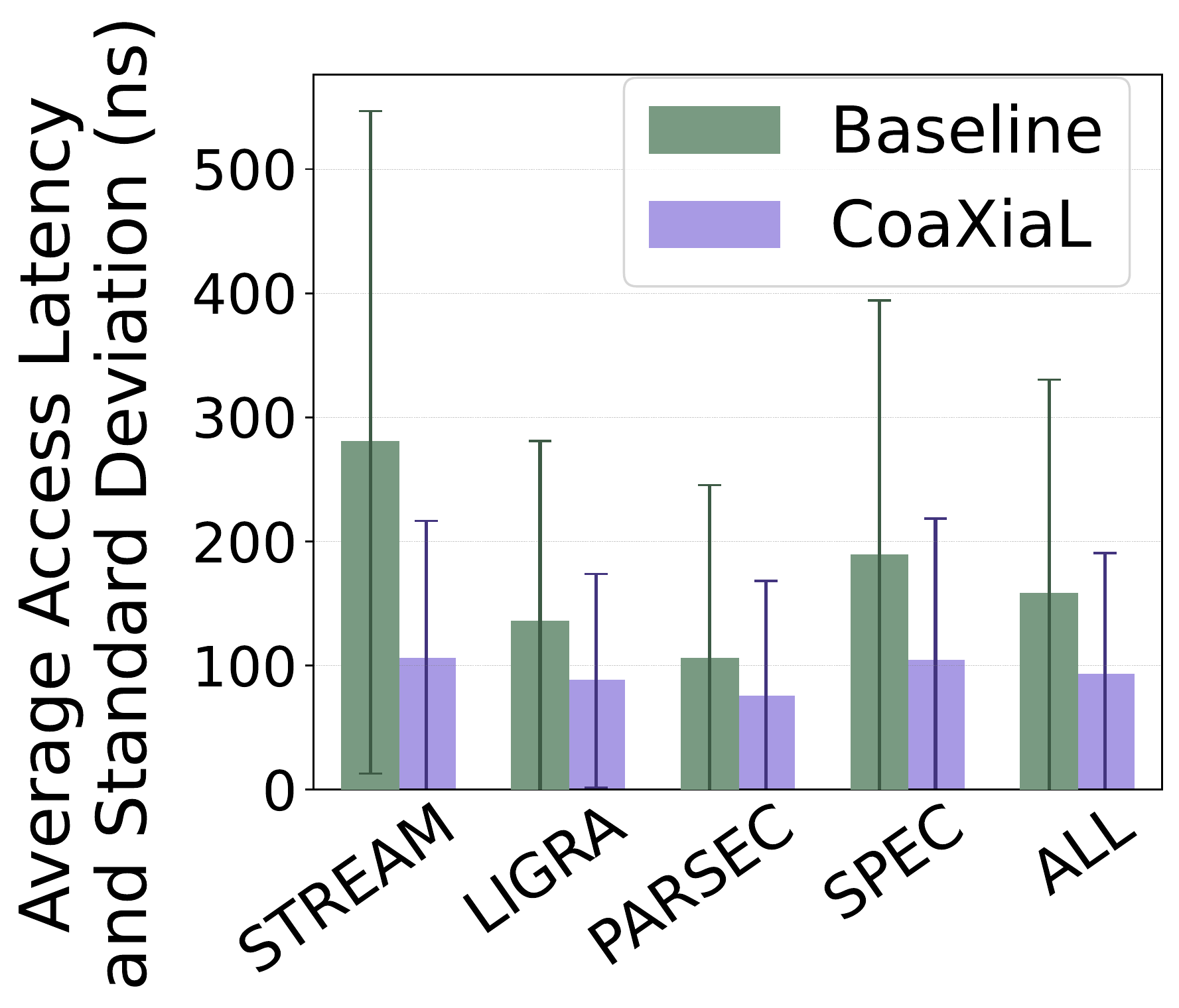} 
    \label{fig:stdev}
    }    
    \hspace{3mm}
    \subfloat[Cumulative Distribution Function (CDF) of memory access time for Streamcluster.]{
    \includegraphics[width=.4\columnwidth]{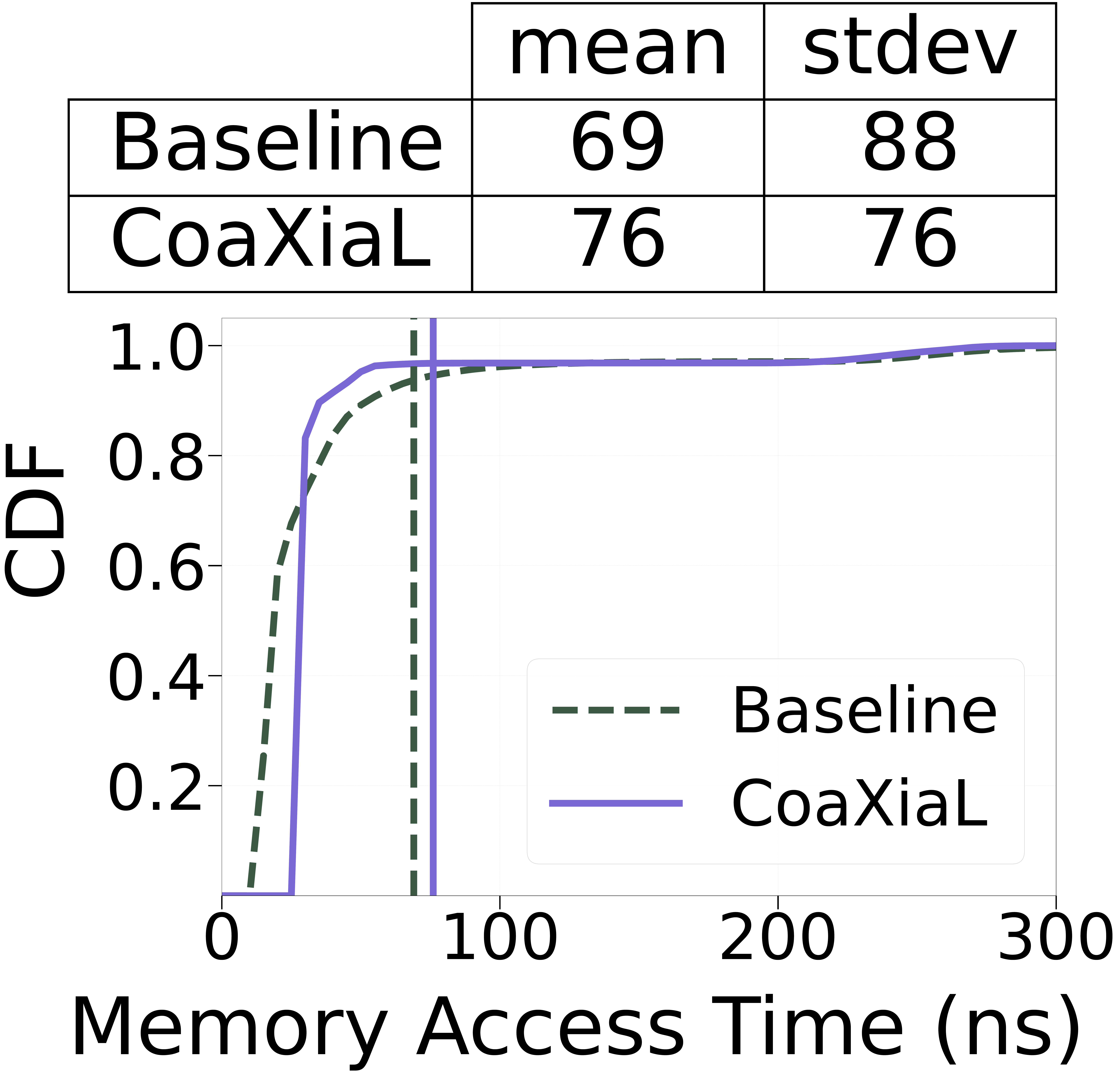} 
    \label{fig:cdf-streamcluster}
    }
    
    \vspace{-2mm}
    
    \caption{Memory access latency distribution.}

\end{figure}

While most of \TheName's performance gains can be justified by the achieved reduction in average memory latency,  a compounding positive effect is reduction in latency variance as evidenced in \cref{sec:motivation:variance}.
For each of the four evaluated workload groups, \cref{fig:stdev} shows the mean average latency and standard deviation (stdev) for \TheName and the DDR-based baseline.
As already seen in \cref{sec:eval:all_apps}, \TheName delivers a 45--60\% reduction to average memory access latency.
\cref{fig:stdev} shows that \TheName also achieves a similar reduction in stdev, indicating lower dispersion and fewer extreme high-latency values.

To further demonstrate the impact of access latency distribution and temporal effects, we study a few workloads in more depth.
\textit{Streamcluster} presents an interesting case because its performance improves despite a slightly higher average memory access latency of 76ns compared to the baseline's 69ns (see \cref{fig:main-eval}).
\cref{fig:cdf-streamcluster} shows the Cumulative Distribution Function (CDF) of Streamcluster's memory access latencies, illustrating that the baseline results in a higher variance than \TheName (stdev of 88 versus 76), due to imbalanced queuing across DRAM banks. The tighter distribution of memory access latency allows \TheName to outperform the baseline despite a 10\% higher average memory access latency.

Some workloads benefit from \TheName more than other workloads with similar or higher memory bandwidth utilization (\cref{fig:main-eval} (bottom)). 
For example, \textit{bwaves} uses a mere 32\% of the baseline's available bandwidth but suffers an overwhelming 390ns queuing delay. 
Even though \textit{bwaves} uses less bandwidth on average compared to workloads (e.g., \textit{radii} with 65\% bandwidth utilization), it exhibits bursty behavior that incurs queuing spikes which can be more effectively absorbed by \TheName.
\textit{Kmeans} exhibits the opposite case. Despite having the highest bandwidth utilization in the baseline system, it experiences a relatively low average queuing delay of 50ns and exhibits one of the lowest latency variance values across workloads, indicating an even distribution of accesses over time and across DRAM banks. 
\textit{Kmeans} is also an outlier with near-zero write traffic, thus avoiding the turnaround overhead from the memory controller switching between read and write mode that results in bandwidth underutilization. %

\begin{figure*}
    \centering

    \includegraphics[width=\linewidth]{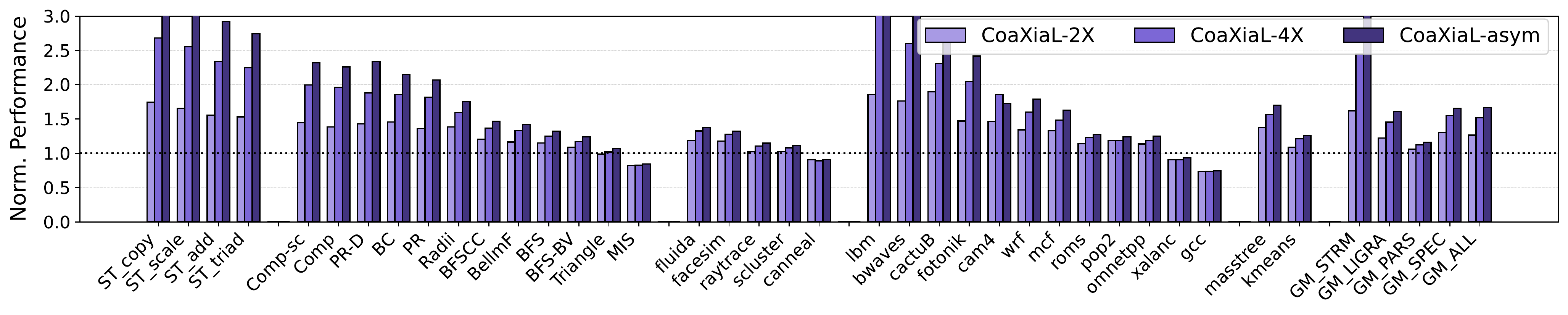}
    
    \vspace{-2mm}
    
    \caption{\TheName's performance at different design points, norm. to the DDR-based server baseline. \TheName-$4\times$ outperforms \TheName-$2\times$, despite its halved LLC size. \TheName-asym considerably outperforms our default \TheName-$4\times$ design.}
    \label{fig:bw-sensitivity}
    
\end{figure*}

\subsection{Alternative \TheName designs}
\label{sec:eval:bw-boost-sensitivity}

\cref{fig:bw-sensitivity} evaluates the two alternative \TheName designs introduced in \cref{sec:design}---\TheName-$2\times$ and \TheName-asym---in addition to our default \TheName-$4\times$. 
\TheName-$2\times$ achieves a $1.26\times$ average speedup over the baseline, down from \TheName-$4\times$'s $1.52\times$ gain. 
This confirms our intuition that doubling memory bandwidth availability at the cost of halving the LLC is beneficial for virtually all workloads.  
\TheName-asym improves performance by $1.67\times$ on average---a considerable 15\% gain on top of \TheName-$4\times$---and no workload is negatively affected by \TheName-asym's reduced write bandwidth.
This result implies an exciting opportunity to improve bandwidth efficiency in memory devices attached via serial interconnects by provisioning the interfaces in a manner that is aware of the workloads' read versus write demands.

\vspace{0.05in} 
\stepcounter{takeaway}
\begin{tcolorbox}[colframe=blue!75!black,colback=white,boxrule=1pt,left=5pt,right=5pt,top=3pt,bottom=3pt]
\textbf{Takeaway \#\thetakeaway{}:} 
Provisioning the lanes in read/write demand aware manner considerably improves performance compared to the default 1:1 read:write provisioning ratio.
\end{tcolorbox}

\subsection{Sensitivity to CXL's Latency Overhead}
\label{sec:eval:lat-sensitivity}

While we base our main evaluation on a 30ns roundtrip CXL interface latency off the CXL 3.0 specification and current industry expectations (see \cref{sec:motivation:cxl-lat}), we also evaluate a more pessimistic latency overhead of 50ns, in case early products do not meet the 30ns target.
Such latency may also better represent CXL-attached memory devices located at a longer physical distance from the CPU, or devices with an additional multiplexing overhead (e.g., memory devices shared by multiple servers---a scenario CXL intends to enable \cite{gouk:direct,li:pond}).

\begin{figure*}
    \centering

    \includegraphics[width=\columnwidth*2]{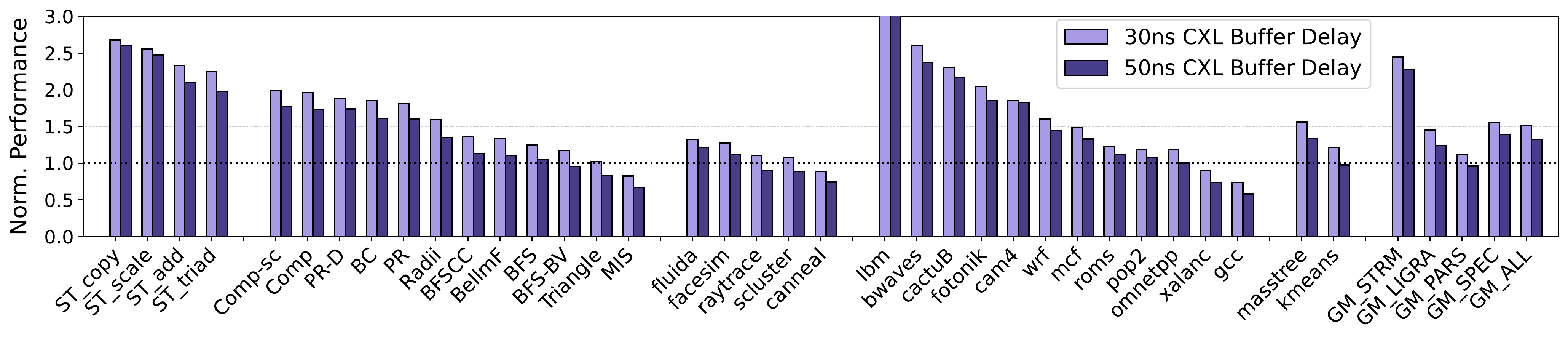}
    
    \vspace{-3mm}
    
    \caption{\TheName's performance for different CXL latency premium, norm. to the DDR-based server. Even with a 50ns interface latency penalty, \TheName yields a {$1.33\times$} average speedup.}
    \label{fig:lat-sensitivity}
    
\end{figure*}

\cref{fig:lat-sensitivity} shows \TheName's performance at 30ns (our default) and 50ns  CXL interface latency overhead, normalized to the DDR-based baseline.
Although increasing latency overhead to 50ns reduces \TheName's average speedup, it remains significant at $1.33\times$.
Memory-intensive workloads continue to enjoy drastic speedups of over 50\%, but more workloads (nine, up from four with 30ns latency penalty) take a performance hit.
These results imply that while a \TheName with a higher CXL latency is still worth pursuing, it should be used selectively for memory-intensive workloads.
Deploying different classes of servers for different optimization goals is common practice not only in public clouds \cite{google-machine-types} but also in private clouds (e.g., different web and backend server configurations) \cite{yosemite, hazelwood:applied}.

\vspace{0.05in} 
\stepcounter{takeaway}
\begin{tcolorbox}[colframe=blue!75!black,colback=white,boxrule=1pt,left=5pt,right=5pt,top=3pt,bottom=3pt]
\textbf{Takeaway \#\thetakeaway{}:} 
Even with a 50ns CXL latency overhead, \TheName achieves a considerable $1.3\times$ average speedup across all workloads.
\end{tcolorbox}

\subsection{Sensitivity to Core Utilization}
\label{sec:eval:unloaded}

\begin{figure*}
    \centering

    \includegraphics[width=\linewidth]{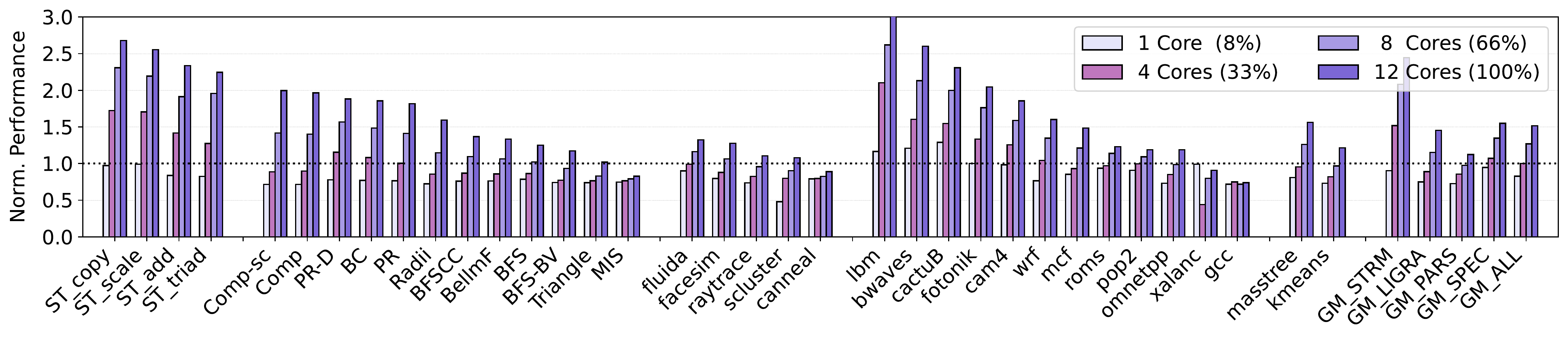}
    
    \vspace{-3mm}
    
    \caption{Performance of \TheName as a function of active cores, norm. to DDR-based server baseline at the same active cores. }
    \label{fig:low-util}
    
\end{figure*}

\cref{fig:low-util} evaluates \TheName's performance under varying levels of system utilization by provisioning proportionately less work on a fraction of cores of the system. %
We first study the extreme case of using a single core on our 12-core simulated system ($8\%$ utilization). In this scenario, virtually all workloads suffer performance degradation with \TheName, for a 17\%  average slowdown. 
\textit{Xalancbmk} exhibits a corner case where the working set fits in the LLC when only one instance is running, removing most memory accesses.
The  extreme single-core experiment showcases \TheName's worst-case behavior, where the memory system is the least utilized.

We then increase the system utilization to 33\% and 66\%, by deploying workload instances on 4 and 8 cores of the 12-core CPU, respectively.
We also show results for 100\% utilization (all cores used) again as a point of comparison. %
\TheName's bandwidth abundance gradually starts paying off, by eliminating the slowdown at 33\% utilization for most workloads, and then delivering significant gains---1.27$\times$ on average and up to 2.62$\times$---even at 66\% utilization. 
The 66\% utilization point can also be considered as a good proxy for a fully loaded system where cores and DDR controllers are provisioned at an 8:1 ratio. %
An 8:1 core:MC ratio is the design point of many server processors with fewer than 100 cores today, such as AMD EPYC Milan and Genoa \cite{anand_milan,amd-genoa}. 
Thus, the $66\%$ utilization results imply that \TheName's approach is applicable beyond high-end throughput-oriented processors that already exhibit 12:1 core:MC oversubscription.

\vspace{0.05in} 
\stepcounter{takeaway}
\begin{tcolorbox}[colframe=blue!75!black,colback=white,boxrule=1pt,left=5pt,right=5pt,top=3pt,bottom=3pt]
\textbf{Takeaway \#\thetakeaway{}:} 
Even at 66\% server utilization---or 8:1 core:MC ratio---\TheName delivers a $1.27\times$ speedup.
\end{tcolorbox}

\subsection{Power Requirements and Energy Efficiency}
\label{sec:eval:power}

Although \TheName's added serial links and $4\times$ more DIMMs increase the server's power consumption, our system also affords much higher throughput. 
To take this power increase into account, we compute the \textit{Energy Delay Product (EDP = system power $\times$ CPI$^2$)} of the baseline and \TheName-4$\times$.
A lower EDP value indicates a more efficient system that consumes less energy to  complete the same work, even if it operates at a higher power.

We model power for a manycore processor similar to AMD EPYC Bergamo (128 cores)~\cite{wccf_bergamo} or Sierra Forest (144 cores)~\cite{wccf_intel_sierra}. 
The latter is expected to have a 500W TDP, which is in line with current processors (e.g., 96-core AMD EPYC Genoa~\cite{amd-genoa} has a TDP of 360W).
While the memory controller and interface require negligible power compared to the processor, we include them for completeness.
We estimate controller and interface power per DDR5 channel to be 0.5W and 0.6W, respectively \cite{volos:memory}, or 13W in total for a baseline processor with 12 channels. %
Similarly, PCIe 5.0's interface power is $\sim$0.2W per lane \cite{bichan:pcie5}, or 77W for the 384 lanes required to support \TheName's 48 DDR5 channels.

A significant fraction of a large-scale server's power is attributed to memory. 
We use Micron's power calculator tool~\cite{micron:calc} to compute our baseline's and CXL system's DRAM power requirement by taking the observed average memory bandwidth utilization of 52\% for baseline and 21\% for \TheName into account. 
As this tool only computes power up to DDR4-3200MT/s modules, we model a 64GB 2-rank DDR4-3200 DIMM (16GB 2-rank module for CXL) and double the power to obtain power consumption of a 128 GB DDR5 channel (32 GB channel for CXL). 
While \TheName employs $4\times$ more DIMMs than the baseline, its power consumption is only $1.75\times$ higher due to lower memory utilization. 

\cref{table:power_analysis} summarizes the key power components for the baseline and \TheName systems.
The overall system power consumption is 713W for the baseline system and 1.18kW for \TheName, a 66\% increase. Crucially, \TheName massively boosts performance, reducing CPI by 34\%. As a result, \TheName reduces the baseline's EDP by a considerable 28\%. 

\vspace{0.05in} 
\stepcounter{takeaway}
\begin{tcolorbox}[colframe=blue!75!black,colback=white,boxrule=1pt,left=5pt,right=5pt,top=3pt,bottom=3pt]
\textbf{Takeaway \#\thetakeaway{}:} 
In addition to boosting performance, \TheName affords a more efficient system with a 28\% lower energy-delay product.
\end{tcolorbox}

\begin{table}%
  \centering
  \begin{footnotesize}
  \caption{Energy Delay Product (EDP $=$ System power $\times$ CPI$^2$) comparison for target 144-core server. Lower EDP is better.}
  \label{table:power_analysis}
  \setlength{\tabcolsep}{3pt}
  \renewcommand{\arraystretch}{1.15}
  \begin{tabular}{|l||c|c|}
    \hline
    \textbf{EDP Component} & \textbf{Baseline} & \textbf{\TheName}\\ \hline \hline
    Processor Package power & 500W  & 500W \\ \hline
    DDR5 MC \& PHY power (all) & 13W & 52W \\ \hline
    DDR5 DIMM power (static and access) & {200W} & {551W} \\ \hline
    CXL's Interface power (idle and dynamic) & N/A & 77W \\    \hline
    Total system power & 713W & 1,180W \\ \hline \hline
    Average CPI (all workloads) & {2.02}  & {1.33} \\ \hline
    \textbf{EDP (all workloads)} & {\textbf{2,909}} & {\textbf{2,087 (0.72$\times$)}}  \\ \hline
  \end{tabular}
  \end{footnotesize}
\end{table}

\subsection{Evaluation Summary}

CXL-based memory systems hold great promise for manycore server processors.
Replacing DDR with CXL-based memory that offers $4\times$ higher bandwidth at a 30ns  latency premium achieves a $1.52\times$ average speedup across various workloads.
Furthermore, a \TheName-asym design  demonstrates opportunity for additional gain ($1.67\times$ average speedup), assuming a modification to the PCIe standard to allow departure from the rigid 1:1 read:write bandwidth provisioning to allow an asymmetric, workload-aware one. 
Even if \TheName incurs a 50ns latency premium, it promises substantial performance improvement ($1.33\times$ on average).
We show that our benefits stem from reduced memory contention:
by reducing the utilization of available bandwidth resources, \TheName mitigates queuing effects, thus reducing both average memory access latency and its variance.

\section{Related Work}
\label{sec:related}

We discuss recent works investigating CXL-based memory system solutions, prior memory systems leveraging serial interfaces, as well as circuit-level and alternative techniques to improve bandwidth and optimize the memory system.

\smallskip\noindent\textbf{Emerging CXL-based memory systems.} Industry is rapidly adopting CXL and already investigating its deployment in production systems to reap the benefits of memory expansion and memory pooling. 
Microsoft leverages CXL to pool memory across servers, improving utilization and thus reducing cost~\cite{li:pond}.
In the same vein, Gouk et al. \cite{gouk:direct} leverage CXL to prototype a practical instance of disaggregated memory~\cite{lim:disaggregated}.
Aspiring to use CXL as a memory expansion technique that will enable a secondary memory tier of higher capacity than DDR, Meta's recent work optimizes data placement in this new type of two-tier memory hierarchy \cite{marouf:tpp}.
Using an FPGA-based prototype of a CXL type-3 memory device, Ahn et al. evaluate database workloads on a hybrid DDR/CXL memory system and demonstrate minimal performance degradation, suggesting that CXL-based memory expansion is cost-efficient and performant \cite{ahn:enabling}.
Instead of using CXL-attached memory as a memory system extension, our work stands out as the first one to propose CXL-based memory as a complete replacement of DDR-attached memory for server processors handling memory-intensive workloads. 

\smallskip\noindent\textbf{Memory systems leveraging serial interfaces.}
There have been several prior memory system proposals leveraging serial links for high-bandwidth, energy-efficient data transfers. Micron's  HMC was connected to the host over 16 SerDes lanes, delivering up to 160GB/s \cite{pawlowski:hybrid}. IBM's Centaur is a memory capacity expansion solution, where the host uses SerDes to connect to a buffer-on-board, which in turn hosts several DDR channels \cite{centaur}.
FBDIMM \cite{ganesh:fully-buffered} leverages a similar concept to Centaur's buffer-on-board to increase memory bandwidth and capacity. An advanced memory buffer (AMB) acts as a bridge between the processor and the memory modules, connecting to the processor over serial links and featuring an abundance of pins to enable multiple parallel interfaces to DRAM modules. Similar to CXL-attached memory, a key concern with FBDIMM is its increased latency.
Open Memory Interface (OMI) is a recent high-bandwidth memory leveraging serial links, delivering bandwidth comparable to HBM but without HBM's tight capacity limitations \cite{coughlin:higher}. Originally a subset of OpenCAPI, OMI is now part of the CXL Consortium.

Researchers have also proposed memory system architectures making use of high-bandwidth serial interfaces.
In MeSSOS' two-stage memory system, high-bandwidth serial links connect to a high-bandwidth DRAM cache, which is then chained to planar DRAM over DDR~\cite{volos:fat}.
Ham et al.  propose disintegrated memory controllers attached over SerDes, aiming to make the memory system more modular and facilitate supporting heterogeneous memory technologies \cite{ham:disintegrated}.
Alloy  combines parallel and serial interfaces to access memory, maintaining the parallel interfaces for lower-latency memory access \cite{wang:alloy}. Unlike our proposal of fully replacing DDR processor interfaces with CXL for memory-intensive servers, Alloy's approach is closer to the hybrid DDR/CXL memory systems that most ongoing CXL-related research envisions.

\smallskip\noindent\textbf{Circuit-level techniques to boost memory bandwidth.} 
HBM~\cite{kim:hbm} and die-stacked DRAM caches offer an order of magnitude higher bandwidth than planar DRAM, but suffer from limited capacity~\cite{jevdjic:die-stacked,loh:efficiently,qureshi:fundamental}.
BOOM~\cite{yoon:boom} buffers outputs from multiple LPDDR ranks to reduce power and sustain server-level performance, but offers modest gains due to low frequency LPDDR and limited bandwidth improvement.
Chen et al.~\cite{chen:increasing} propose dynamic reallocation of power pins to boost data transfer capability from memory during memory-intensive phases, during which processors are memory bound and hence draw less power.
Pal et al. \cite{pal:case} propose packageless processors to mitigate pin limitations and boost the memory bandwidth that can be routed to the processor.
Unlike these proposals, we focus on conventional processors, packaging, and commodity DRAM, aiming to reshape the memory system of server processors by leveraging the widely adopted up-and-coming CXL interconnect.

\smallskip\noindent\textbf{Other memory system optimizations.} 
Transparent memory compression techniques are a compelling approach to increasing effective memory bandwidth \cite{young:enabling}. 
Malladi et al.~\cite{malladi:towards} leverage mobile LPDDR DRAM devices to design a more energy-efficient memory system for servers without  performance loss.
These works are orthogonal to our proposed approach.
Storage-class memory, like Phase-Change Memory~\cite{fong2017phase} or Intel's Optane \cite{optane},  has attracted significant interest as a way to boost a server's memory capacity, triggering research activity on transforming the memory hierarchy to best accommodate such new memories \cite{agarwal:thermostat, dulloor:data, lee:architecting, ustiugov:design}. 
Unlike our work, such systems often trade off bandwidth for capacity.

\section{Conclusion}
\label{sec:conclusion}

Technological trends motivate a server processor design where all memory is attached to the processor over the emerging CXL interconnect instead of DDR. 
CXL's superior bandwidth per pin helps bandwidth-hungry server processors scale the bandwidth wall.
By distributing memory requests over $4\times$ more memory channels, CXL reduces queueing effects on the memory bus.
Because queuing delay dominates access latency in loaded memory systems, such reduction more than compensates for the interface latency overhead introduced by CXL.
Our evaluation on a diverse range of memory-intensive workloads shows that our proposed \TheName server delivers $1.52\times$ speedup on average, and up to $3\times$. %

\bibliographystyle{IEEEtranS}
\balance
\bibliography{gen-abbrev,dblp,misc}

\end{document}